\documentclass[preprint]{elsarticle}
\usepackage{graphicx}
\usepackage{dcolumn}
\usepackage{bm}
\usepackage{mhchem}
\usepackage[dvips]{color}
\usepackage{url}

\begin{document}

\begin{frontmatter}
\title{Bayesian inference of in-medium baryon-baryon scattering cross sections from HADES proton flow data}

\author{Bao-An Li$^*$\footnote{Corresponding author: Bao-An.Li$@$tamuc.edu}}
\author{Wen-Jie Xie$^{+}$\footnote{wenjiexie@yeah.net}}
\address{$^{*}$Department of Physics and Astronomy, Texas A$\&$M University-Commerce, Commerce, TX 75429-3011, USA}
\address{$^{+}$Department of Physics, Yuncheng University, Yuncheng 044000, China}

\date{\today}
\setcounter{MaxMatrixCols}{10}

\begin{abstract}
Within a Bayesian statistical framework using a Gaussian Process emulator for an isospin-dependent Boltzmann-Uehling-Uhlenbeck (IBUU) transport model simulator of heavy-ion reactions at intermediate energies, we infer from the HADES proton flow data the posterior probability distribution functions of in-medium baryon-baryon scattering cross section modification factor X with respect to free-space and the corresponding incompressibility K of nuclear matter as well as their correlation function. The mean value of X is found to be $X=1.32^{+0.28}_{-0.40}$ at 68\% confidence level  assuming the nuclear incompressibility K will not exceed 400 MeV, providing circumstantial evidence for enhanced baryon-baryon scattering cross sections in hot and dense nuclear matter.

\end{abstract}



\end{frontmatter}
\section{Introduction}\label{S1}
Comparisons of hydrodynamics and/or transport model predictions with experimental data of various components of nuclear collective flow in heavy-ion collisions over a broad energy range have revealed much needed information about the Equation of State (EOS) and transport properties (e.g, viscosity or in-medium nucleon-nucleon (NN) scattering cross sections) of hot and dense nuclear matter \cite{Sto86,Bertsch,Cas90,das93,res97,Bass,Pawel02,Buss,Heinz13,QCD} while many interesting issues remain to be addressed \cite{EOSBook,LRP1,LRP2,LRP3}. In the intermediate beam energy range from around 30 MeV/nucleon to several GeV/nucleon, it is know that there is a degeneracy between the stiffness (normally measured/labeled by using the incompressibility $K$ of symmetric nuclear matter at saturation density $\rho_0$) of nuclear EOS and the in-medium baryon-baryon scattering cross sections ($\sigma^{med}_{NN}$) in describing flow observables, see, e.g., Refs.
\cite{Bertsch:1988xu,Xu:1991zz,Westfall:1993zz,Klakow:1993dj,TLi93,Alm:1995chb,BALI1,Zheng99,LiSustich,Danielewicz:2002he,BALI2}.
Namely, the same flow data can normally be described equally well by either modifying the incompressibility $K$ or the in-medium cross section $\sigma^{med}_{NN}$. While theoretically it has been a longstanding goal to calculate the $\sigma^{med}_{NN}$ and K consistently within the same microscopic many-body theoretical framework using the same nuclear effective interactions \cite{Herman1}, practically most of the information about them obtained so far are from comparing transport model predictions with experimental observations. However, essentially all previous efforts to extract the $\sigma^{med}_{NN}$ and K
from intermediate energy heavy-ion collisions have been using the forward-modeling approach, i.e., comparing directly experimental data with transport model predictions using typically a stiff ($K\approx 380$ MeV) and/or soft ($K\approx 200$ MeV) EOS (or $2\sim 3$ Skyrme EOSs) and sometimes with a cascade option coupled with $1\sim 3$ parameterization(s) for $\sigma^{med}_{NN}$ or simply using the free-space experimental cross sections $\sigma^{free}_{NN}$. Generally speaking, only broad constraining bands or qualitative conclusions have been made from such approach without giving quantified uncertainties for neither the in-medium cross section $\sigma^{med}_{NN}$ nor the stiffness K of nuclear EOS.

With few exceptions (e.g., Ref.  \cite{Zhang:2007gd}) strong indications of a reduced $\sigma^{med}_{NN}$ relative to $\sigma^{free}_{NN}$ were found from analyzing the collective flow and nuclear stopping power of heavy-ion collisions up to beam energies about 800 MeV/nucleon, see, e.g., Refs. \cite{PLi18,Li:2022wvu} and references therein.  At higher beam energies when the inelastic baryon-baryon scatterings dominate, the situation becomes very uncertain.  Theoretically, nuclear medium affects baryon-baryon scattering cross sections through several factors \cite{Haar} including the relative velocity thus the incoming current of two colliding baryons through their effective masses due to the momentum dependence of their mean-field potentials \cite{Gale,LiChen05,Li:2005iba}, the interaction matrix element itself due to, e.g., off-shell or collectivity of exchanged mesons \cite{Bertsch:1988xu,Li:1993rwa,Lom96,Fuchs01}, and the Pauli blocking of both intermediate and final states \cite{Sa,Carlos}. There are unresolved fundamental issues about each of these elements. For example, considering the fact that nucleon effective masses are reduced compared to their free-space values, many transport model simulations have adopted reduced in-medium cross sections based on the effective mass scaling by assuming that the baryon-baryon interaction matrix element is the same as in free-space, see, e.g., Ref \cite{BALI} for a review. On the other hand, considering the pion collectivity in dense matter in evaluating the in-medium interaction matrix element, inelastic nucleon-nucleon cross sections in dense medium were found to increase significantly with respect to their free-space values \cite{Bertsch:1988xu}. Moreover, there are many interesting technical issues in modeling scatterings in dense medium within transport models, see, e.g., Ref. \cite{Herman} for a recent review. Thus, how to extract the $\sigma^{med}_{NN}$ effectively, and most importantly what is its value with quantified uncertainties, from experimental observables remains a major challenge in intermediate energy heavy-ion physics. Besides its importance for understanding fundamental physics mentioned above, reliable knowledge about the $\sigma^{med}_{NN}$ has important ramifications for space radiation protection, nuclear stockpile stewardship, and nuclear waste transmutation.

In this work, a Bayesian inference of the in-medium baryon-baryon scattering modification factor $X\equiv\sigma^{med}_{NN}/\sigma^{free}_{NN}$ from HADES proton flow data \cite{Hades,Ha2} is carried out with a uniform prior for $X=0.5\sim 2$ and $K=180\sim 400$ MeV, respectively, using a Gaussian Process (GP) emulator for the isospin-dependent Boltzmann-Uehling-Uhlenbeck (IBUU) transport model simulator \cite{BALI1,BALI,libauer2,LiBA04}. Circumstantial evidence for an enhanced in-medium baryon-baryon scattering cross section is found with a mean value of $X=1.32^{+0.28}_{-0.40}$ and $K=346^{+29}_{-31}$ MeV at 68\% confidence level from their posterior probability distribution functions (PDFs) inferred.

\section{A brief summary of the IBUU simulator used in this work}
The IBUU simulator has multiple choices for the baryon mean-field potential and in-medium baryon-baryon cross sections \cite{BALI}. For the purpose of this work, considering the limitations of our computing powers and the task to generate the necessarily large data sets to train and test the GP emulator of the IBUU simulator, we choose the Skyrme-type momentum-independent potential for baryon $q$
\begin{equation}
      V_{q}(\rho,\delta) = a (\rho/\rho_0) + b (\rho/\rho_0)^{\sigma}\
	+V_{\rm asy}^{q}(\rho,\delta) +V^{q}_{\rm Coulomb}.
\end{equation}
The parameters $a,~b$ and $\sigma$ are determined by the
saturation properties and the compressibility $K$ of symmetric nuclear
matter at $\rho_0$ by \cite{Bertsch,BALI1}
\begin{eqnarray}
a&=&-29.81-46.90\frac{K+44.73}{K-166.32}~({\rm MeV}),\\
b&=&23.45\frac{K+255.78}{K-166.32}~({\rm MeV}),\\
\sigma&=&\frac{K+44.73}{211.05}.
\end{eqnarray}
To avoid causing confusions, we emphasize a few points here. Firstly, the above Skyrme potential has only one free parameter $K=9\rho_0^2[d^2E_0(\rho)/d \rho^2]_{\rho_0}$
that determines not only the incompressibility of SNM at $\rho_0$ but also its high-density stiffness measured by the skewness parameter $J_0=27\rho_0^3[d^3E_0(\rho)/d \rho^3]_{\rho_0}$ and the kurtosis parameter $Z_0=81\rho_0^4[d^4E_0(\rho)/d \rho^4]_{\rho_0}$ where $E_0(\rho)$ is the energy per nucleon in SNM at density $\rho$. Analytically, one can easily show that once the K parameter is specified for a Skyrme energy density functional $E_0(\rho)$ both the $J_0$ and $Z_0$ are then uniquely determined. Thus, the stiffness of Skyrme EOS in the entire density range is completely determined by the single parameter K. Nevertheless, we caution that nuclear collective flow in heavy-ion collisions at beam energies around 1-2 GeV/nucleon depend individually not only on the incompressibility K of SNM at $\rho_0$ but also the high density stiffness of the EOS. More specifically, the pressure in cold SNM is well described up to about $(4-5)\rho_0$ by \cite{Xie20}
 \begin{equation}\label{pressure0}
  P(\rho)=\rho^2\frac{dE_0(\rho)}{d\rho}=\frac{\rho^2}{\rho-\rho_0}[K(\frac{\rho-\rho_0}{3\rho_0})^{2}+\frac{J_0}{2}(\frac{\rho-\rho_0}{3\rho_0})^{3}+\frac{Z_{0}}{6}(\frac{\rho-\rho_0}{3\rho_0})^4].
\end{equation}
In such meta-model EOS, the three parameters K, $J_0$ and $Z_0$ are independent {\it a priori}.
A recent Bayesian calibration of the above pressure by us \cite{Xie20} using the constraining band on pressure in SNM in the density region of $(1.2\sim 4.5)\rho_0$ from analyzing the kaon production and the beam energy dependence of nuclear collective flow in relativistic heavy-ion collisions \cite{Pawel02,Fuchs,Lynch09} indicates that the pressure underlying the collective flow constrain not only the K but also both the $J_0$ and $Z_0$ when they were used as three independent parameters. Moreover, the constraining band of pressure from analyzing flow in relativistic heavy-ion collisions makes the posterior PDFs of both $J_0$ and $Z_0$ much narrower than their uniform prior PDFs. Nevertheless, the posterior PDFs of $J_0$ and $Z_0$ depend strongly on the prior range and PDF of K. And, of course, the posterior PDFs of the three parameters are all correlated as one expects. Therefore, while the single parameter K is enough to label accurately the entire Skyrme EOS, the flow data to be used in our analyses constrain not only the stiffness of SNM at $\rho_0$ but also at higher densities. How the pressure in each density region contributes to the flow is consistently determined by the BUU reaction dynamics through the Skyrme mean-field potential given above.

It is well known that the momentum-dependence of nuclear mean-field affects the nuclear collective flow, see, e.g., Refs. \cite{Aich87,Gale87,Prakash,Gerd,Pan-Pawel,Cozma}. Nevertheless, essentially all dynamical effects due to the momentum dependence can be  mimicked by varying the incompressibility $K$ within a momentum-independent model, i.e., the same flow data and generally all observables in heavy-ion collisions can be equally well reproduced by using a smaller K with momentum dependence or a larger K without it, see, e.g., Ref. \cite{Fuchs06} for a review. As we are going to generate the value of K randomly with an equal probability within its prior range of 180 MeV to 400 MeV in our Bayesian analysis using the above momentum-independent potential, our choice for the mean-field potential is physically sufficient for the present study. The $V_{\rm asy}^{q}(\rho,\delta)$ is the baryon symmetry potential in asymmetric nuclear matter with isospin asymmetry $\delta$. Here we adopt (corresponding to the $F_3$ in Eq. (3) of Ref. \cite{Li97}) the following momentum-independent symmetry potential
\begin{equation}
V_{\rm asy}^{n(p)}=\pm 2e_a (\rho/\rho_0)^{1/2}\delta-\frac{1}{2}e_a(\rho/\rho_0)^{1/2}\delta^2
\end{equation}
where $e_a\equiv E_{\rm sym}(\rho_0)-(2^{2/3}-1)\,{\textstyle\frac{3}{5}}E_F^0$ with the symmetry energy $E_{\rm sym}(\rho_0)=32$ MeV and the Fermi energy $E_F^0=36$ MeV at $\rho_0$. The $\pm$ sign is for neutrons/protons, respectively, and $V^{q}_{\rm Coulomb}$ is the Coulomb potential for charged particles. The potentials of baryon resonances ($\Delta$ and $N^*$) are related to those of nucleons through the square of the Clebsch-Gordon coefficients in their decays to pion+nucleon processes \cite{Linpa}.

We notice that the above choice of the single-nucleon potential is still among the most widely used options, see, e.g., Ref. \cite{Ste}. Instead of introducing more parameters to distinguish possibly different in-medium effects for elastic and inelastic scatterings and their energy dependence, in this work we use the single parameter X to modify all free-space experimental nucleon-nucleon scattering cross sections used as default in the original IBUU code. Nevertheless, considering the current situation of the field, we purposely use the very broad prior ranges for both the parameter X and K. As we shall show in the next section, within the Bayesian framework, just two data points alone from the proton directed and elliptical flow at a single beam energy can already narrow down significantly the PDFs of X and K. While our Bayesian analyses in this work use only two free parameters (X and K) and two data points, the framework established here can be easily extended to more model parameters and use more data in the future. In particular, a Gogny-type flexible parameterization for the single-nucleon potential with momentum dependence for both its isoscalar and isovector parts leading to varying density dependence of nuclear symmetry energy together with a coalescence model for forming light clusters will be carried out in a future work. It is also interesting to note here that several groups are carrying out both forward-modelings and Bayesian analyses with similar goals and some using the same data as ours but employing different dynamical models \cite{Rei,Oma,Oli}. Nevertheless, to our best knowledge, our work here is the only one focusing on inferring the in-medium baryon-baryon cross sections.

\section{IBUU simulations of heavy-ion reactions on the X-K Latin Hyperlattice for training and testing the Gaussian Process emulator}
Since the pioneering work of Refs. \cite{pawel85,oll,art}, the first ($v_1$) and second ($v_2$) coefficients of the Fourier decomposition of particle azimuthal angle distribution $\frac{2\pi}{N}\frac{dN}{d\phi} = 1 + 2\sum_{n=1}^{\infty}v_n\cos{[n(\phi)]}$
have been used to measure the strength of the so-called directed (transverse) and elliptical flow, respectively.
Their values at rapidity $y$ and transverse momentum $p_t$ can be evaluated from
$v_1(y,p_t)=\left<cos(\phi)\right>(y,p_t)=\frac{1}{n}\sum_{i=1}^{n}\frac{p_{ix}}{p_{it}}$ and
$v_2(y,p_{t})=\left<cos(2\phi)\right>(y,p_t)=\frac{1}{n}\sum_{i=1}^{n}\frac{p_{ix}^2-p_{iy}^2}{p_{it}^2}$, where $p_{ix}$ and $p_{iy}$ are the x- and y-component of the $i^{{\rm th}}$ particle momentum, respectively. In our setup of the simulations, the reaction plane is in the $x-o-z$ plane. In the present study, we only use the slope $F_1\equiv dv_1/dy'|_{y'=0}=0.46\pm 0.03$ of proton directed flow $v_1$ at mid-rapidity $y'\equiv y_{\rm cm}/y_{\rm mid}=0$ in the center of mass (cm) frame of the two colliding nuclei, and the proton elliptical flow $v_2=-0.06\pm 0.01$ data for $|y_{cm}|\leq 0.05$ and $p_t\geq 0.3$ GeV/c in mid-central (10-30\% centrality) Au+Au collisions at $E_{\rm beam}/A$=1.23 GeV from the HADES Collaboration \cite{Hades,Ha2}. For this reaction, $y_{\rm mid}=0.74$. The $F_1$ and $v_2$ for this reaction are consistent with those from the FOPI Collaboration \cite{FOPI}. Interestingly, they are respectively at the top and bottom of the $F_1$ vs beam energy and $v_2$ vs beam energy systematic curves based on data accumulated over the last 40 years \cite{EOSBook,Ha2}, indicating that the strengths of both direct flow (positive) and elliptic flow (negative) are the strongest around this beam energy. These data may thus provide the best opportunity for determining the underlying agents creating the collective flow.

We identify free nucleons as those with local densities less than $\rho_0/8$ in the final state of the reaction. In generating the IBUU training and testing data, for each set of the X and K parameters we use 200 testparticles/nucleon and generate randomly 100 impact parameters $b$ with the probability density $P(b)\propto b$ between b=6 fm and 9 fm corresponding to approximately the (10-30)\% centrality of the Au+Au reaction \cite{Ha3}. Thus, for each X and K parameter set, 20,000 IBUU events of Au+Au collisions are used in calculating the $F_1$ and $v_2$ in each step of the Markov Chain Monte Carlo (MCMC) in our Bayesian analyses.
\begin{figure*}[thb]
\centering
\includegraphics[width=0.495\textwidth]{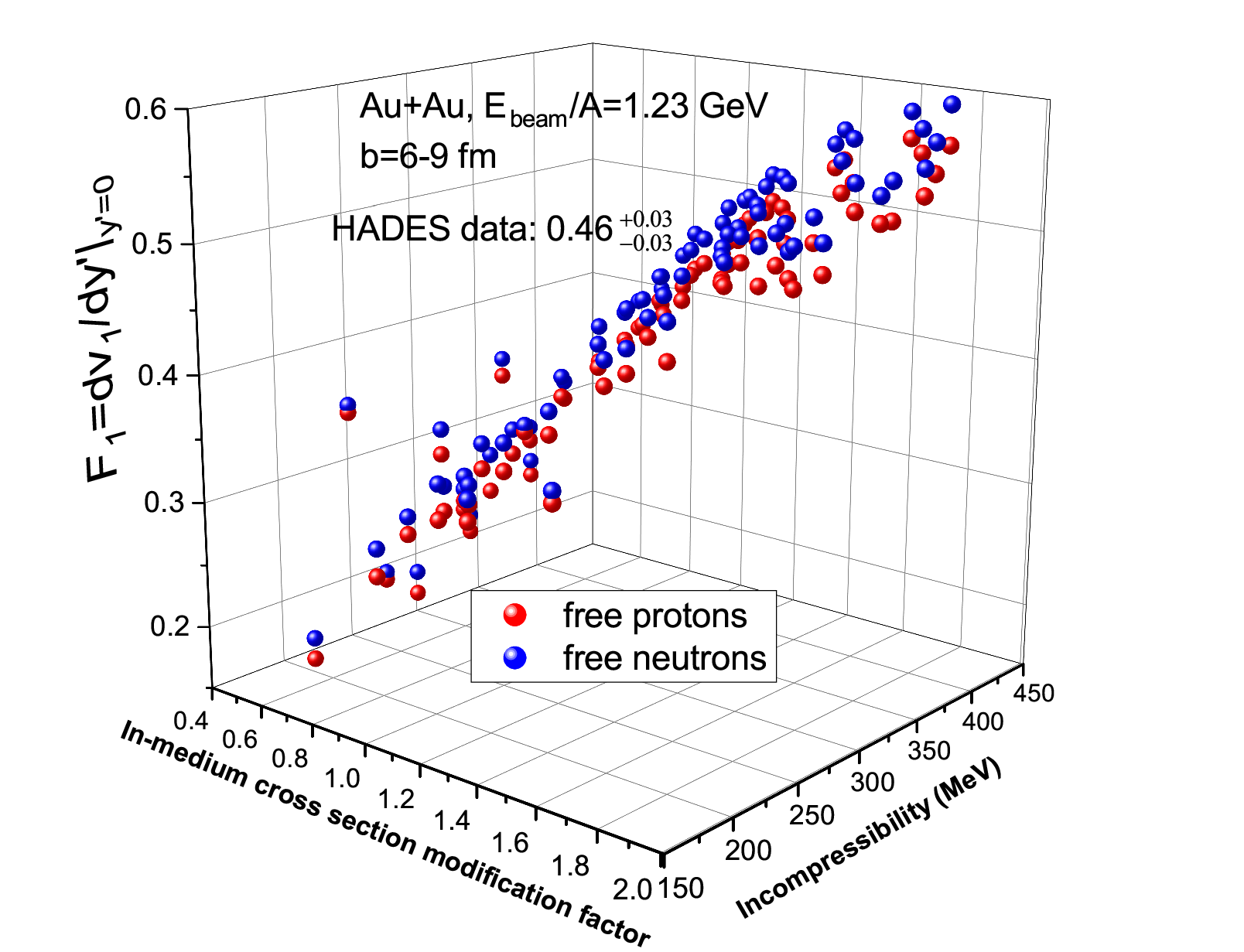}
\includegraphics[width=0.495\textwidth]{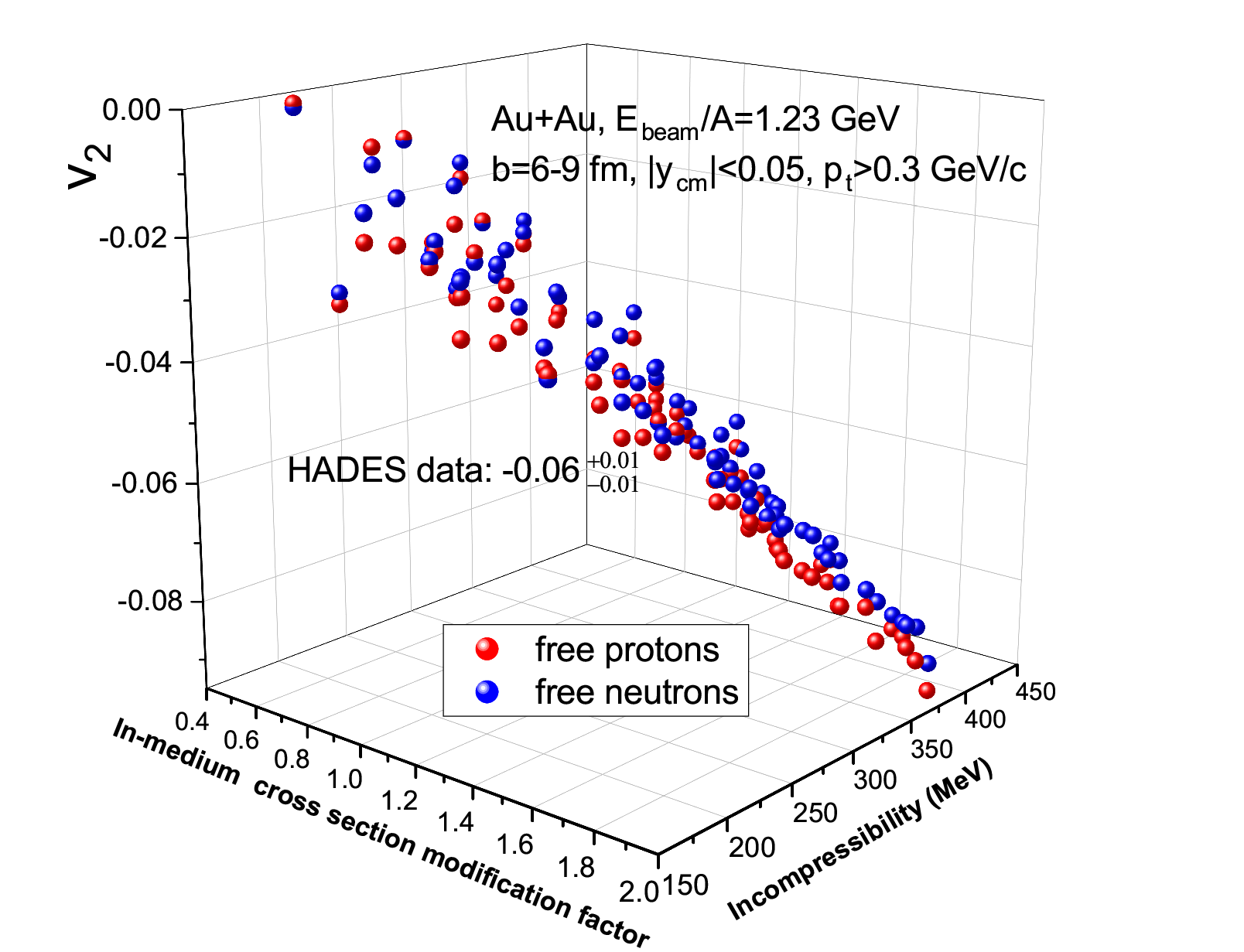}
\caption{Predicted slope $F_1$ of the free proton (red) and neutron (blue) direct flow at mid-rapidity $dv_1/dy'|_{y'=0}$ (left) and elliptical flow $v_2$ (right), respectively, as functions of the incompressibility $K$ (right axis) and in-medium baryon-baryon scattering cross section modification factor X (left axis) generated on a Latin Hyperlattice in the X-K plane using the IBUU transport model for the indicated Au+Au reactions measured by the HADES Collaboration.} \label{v12}
\end{figure*}

Since multi-million MCMC steps are needed in the Bayesian analyses, a fast emulator for the IBUU simulator is necessary. We use here the widely used GP emulator with the Squared Exponential (SE) covariance function \cite{GP}. To train and test it, large sets of IBUU simulations have to be carried out first. To ensure that the emulator can explore unbiasedly the whole X-K parameter plane, we train it on the Latin Hyperlattice \cite{LH} with totally 80 sets of X-K parameters. Shown in Fig. \ref{v12} are the IBUU
simulator predicted slope $F_1$ of the free proton (red) and neutron (blue) directed flow at mid-rapidity $F_1=dv_1/dy'|_{y'=0}$ (left) and elliptical flow $v_2$ (right), respectively, as functions of $K$ (right axis) and  X (left axis) for the indicated Au+Au reactions measured by the HADES Collaboration \cite{Hades,Ha2}. Some interesting observations can be made qualitatively. Firstly, the magnitudes of both $F_1$ and $v_2$ increase with both K and X, thus for a given data set of $F_1$ and $v_2$ the required values of X and K should be anti-correlated. Secondly, the HADES data prefer a very stiff EOS with $K\approx 350$ MeV but also an enhanced in-medium cross section with $X\approx 1.3$.
Thirdly, there are only small differences between the results for protons and neutrons. Nevertheless, it will be interesting to study in the future how these differences may depend on the baryon symmetry potential used and if they can be used to probe the density dependence of nuclear symmetry energy especially at suprasaturation densities.

We notice that there are some outliers from the general trends mostly in the region where both the EOS is soft and the in-medium cross section is reduced. In these cases, there are only few free nucleons emitted and the resulting $F_1$ and $v_2$ values are very small. In their automated calculations in our codes using the same number of test particles within the same rapidity and/or transverse momentum bins, some outliers are expected. Nevertheless, since these $F_1$ and $v_2$ values generated using unrealistically small K and X are so far away from the experimental data, the likelihood function in our Bayesian analyses enables the MCMC process to step away quickly from these points, thus they do not really affect our final conclusions.

\begin{figure*}[thb]
\centering
\includegraphics[width=0.7\textwidth]{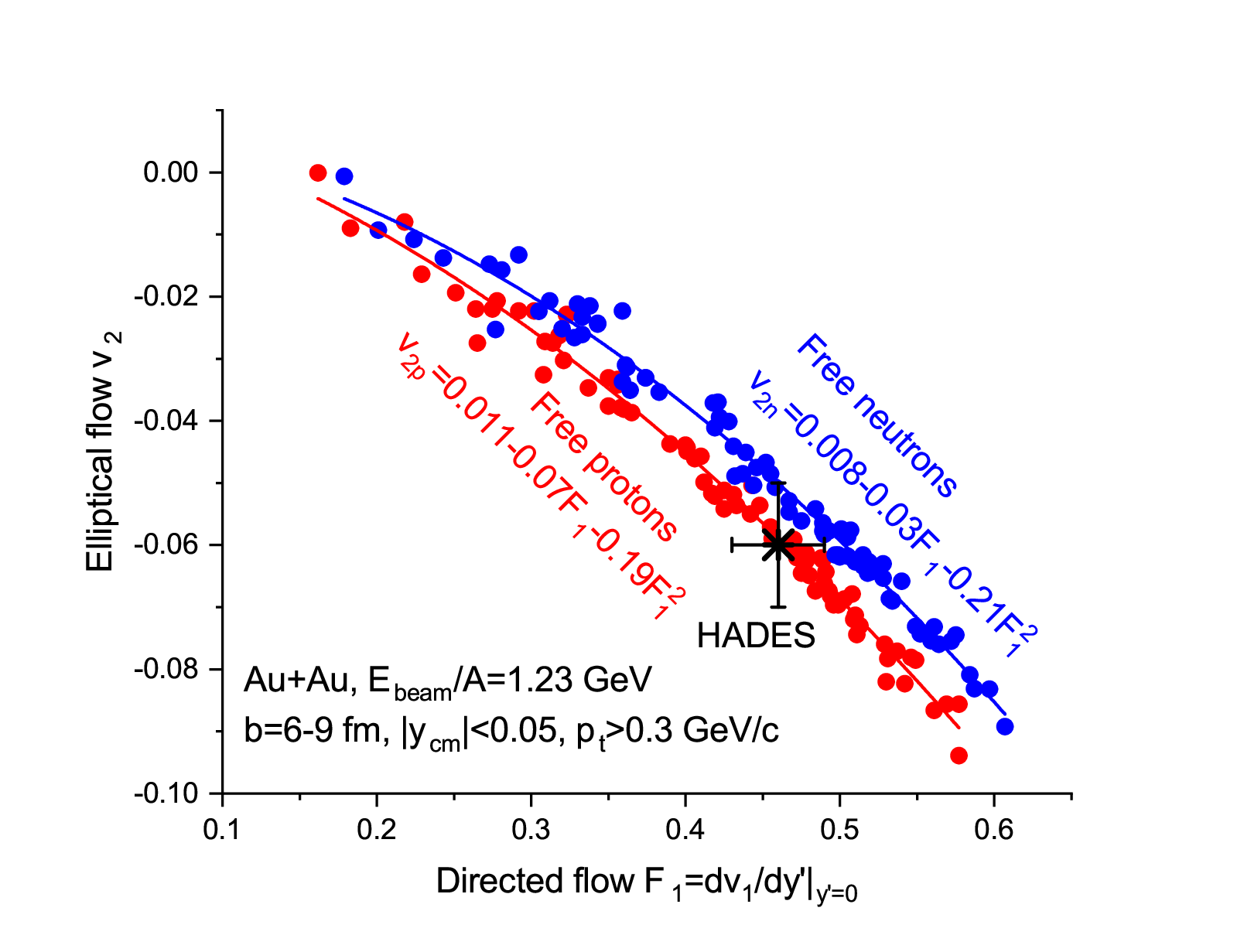}
\caption{Correlation between $v_2$ and $F_1$ for protons and neutrons, separately, for the same results shown in Fig. \ref{v12}. } \label{c12}
\end{figure*}
To see more clearly the correlation between the two observables used here, shown in Fig. \ref{c12} are $v_2$ vs $F_1$ for protons and neutrons, separately. The experimental data from HADES is shown with the black star symbol. We emphasize that the $v_1$ and $v_2$ measure nucleon collective behaviors in different kinematic regions, they thus provide complementary information. Together they constrain more tightly the model parameters although they are strongly correlated. As shown in detail in Ref. \cite{Li-Jake}, in the traditional analyses of the differential flow strengths $v_1(y,p_t)$ and $v_2(y,p_t)$ in different rapidity intervals for the same reaction studied here within the IBUU model, the integrated $v_1(y)$ is dominated by contributions from high $p_t$ particles around both mid-rapidity and projectile/target rapidities. While there are only few high $p_t$ particles compared to the mostly isotropic low $p_t$ particles, the large $p_x$ of these high $p_t$ particles make major contributions to the in-plane flow $v_1(y)$. Due to the total momentum conservation in the transverse directions, the integrated $v_2$ at mid-rapidity for high $p_t$ particles would become more negative as the $F_1$ increases with the increasing X and/or K.
This feature is clearly demonstrated in Fig. \ref{c12}. Moreover, the difference between the $v_2$ vs $F_1$ correlations for protons and neutrons is appreciable.

\begin{figure}[thb]
\centering
\includegraphics[width=0.49\textwidth]{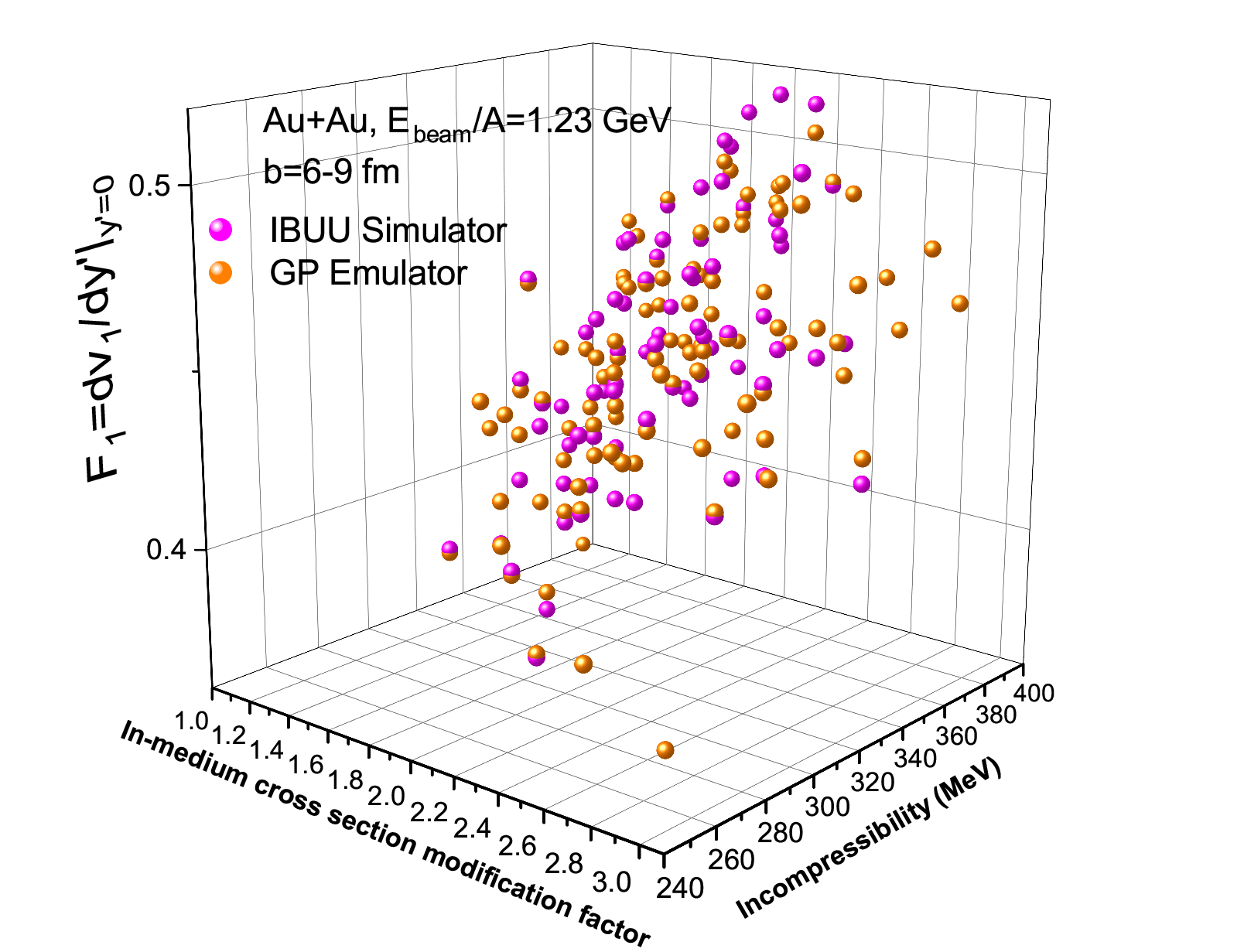}
\includegraphics[width=0.49\textwidth]{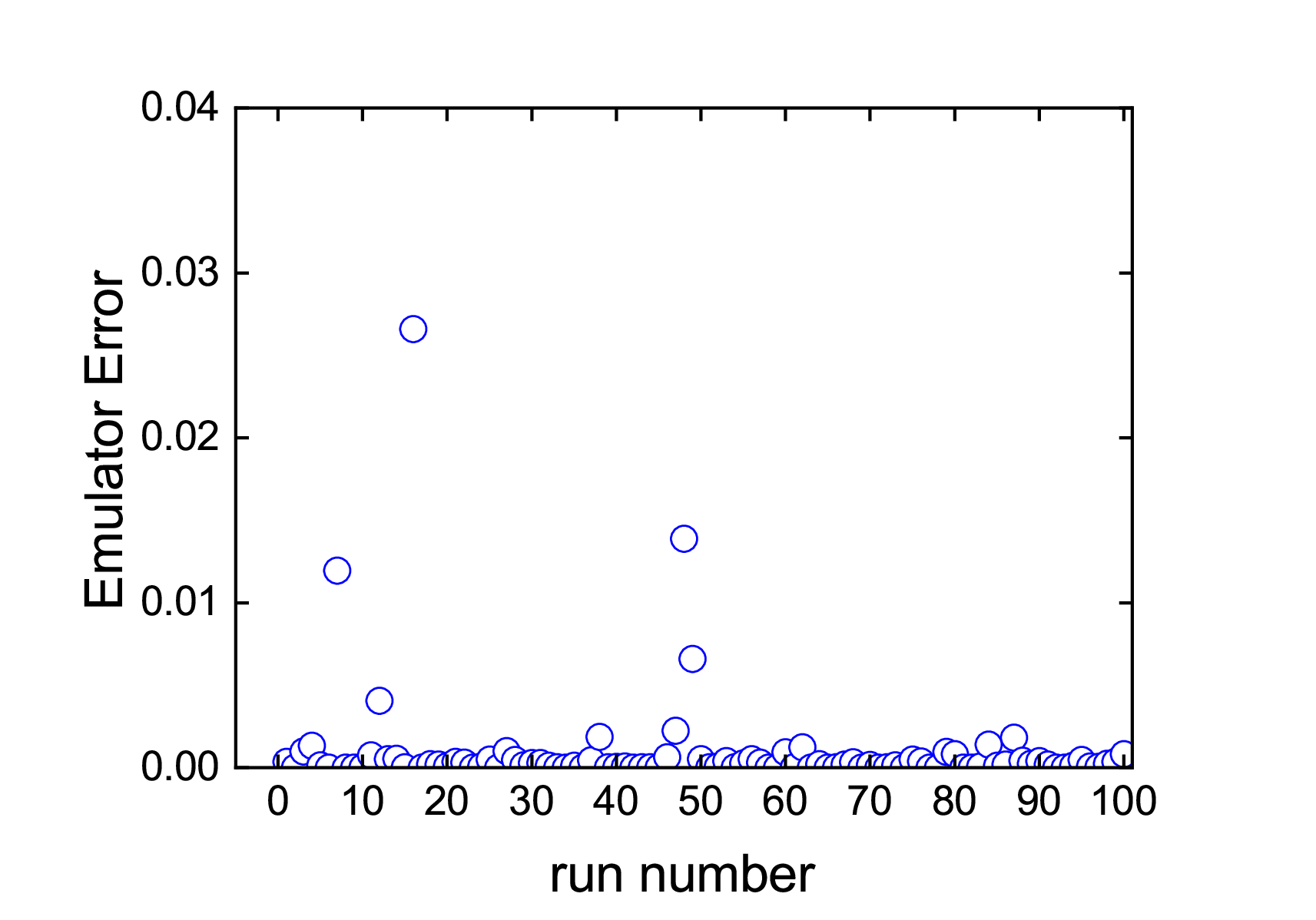}
\caption{Left: A comparison between predictions for $F_1$ by the IBUU simulator and its GP emulator for 100 sets (runs) of X and K parameters randomly generated. Right: the emulator error (EE) as a function of run number. }\label{test}
\end{figure}
Before using the GP emulator in the MCMC, we have tested its reliability using several approaches normally used in the literature \cite{Pratt,Scott,Bass2,Bass3,Ohio} by generating independent training and testing data sets from running the IBUU simulator. By design, the GP emulator pass through the training points. Its power in predicting reliably observables at random locations away from the training points can be evaluated by examining, for instance, the Emulator Error (EE), by comparing predictions of the emulator (emu) and the simulator (sim). In our case, the EE is given by
\begin{equation}
{\rm EE}({\rm run})=[F_1^{{\rm emu}}({\rm run})-F_1^{{\rm sim}}({\rm run})]^2+[v_2^{{\rm emu}}({\rm run})-v_2^{{\rm sim}}({\rm run})]^2
\end{equation}
for each simulator run randomly in the model parameter space. For example, shown in the left panel of Fig. \ref{test} is a direct comparison
between predictions for $F_1$ from the GP emulator and the IBUU simulator with 100 sets (runs) of X and K parameters randomly generated in the regions shown. The corresponding EE is shown as a function of run number in the right panel.  Except 5 runs with very small EE vales less than 0.03, all other runs have EE values very close to zero, indicating that the emulator is working faithfully.

\section{Bayesian inference of the posterior PDFs of X and K}
For completeness, we first recall here the Bayesian theorem
\begin{equation}\label{Bay1}
P({\cal M}|D) = \frac{P(D|{\cal M}) P({\cal M})}{\int P(D|{\cal M}) P({\cal M})d\cal M},
\end{equation}
where the denominator is a normalization constant. The $P({\cal M}|D)$ represents the posterior PDF of the model $\cal M$ given the data set $D$. The $P(D|{\cal M})$ is the likelihood function obtained by comparing predictions of the model $\cal M$ with the data $D$, while the $P({\cal M})$ is the prior PDF of the model $\cal M$. In the present study,
M represents the model parameters X and K, and D represents the $F_1$ and $v_2$ data from HADES. We use the standard Gaussian likelihood function and a constant prior PDF. We note that the training and Bayesian analyses are all done so far in exactly the same ranges of $X = 0.5\sim 2$ and $K = 180\sim 400$ MeV.
\begin{figure}[thb]
\centering
\includegraphics[width=1.5\textwidth]{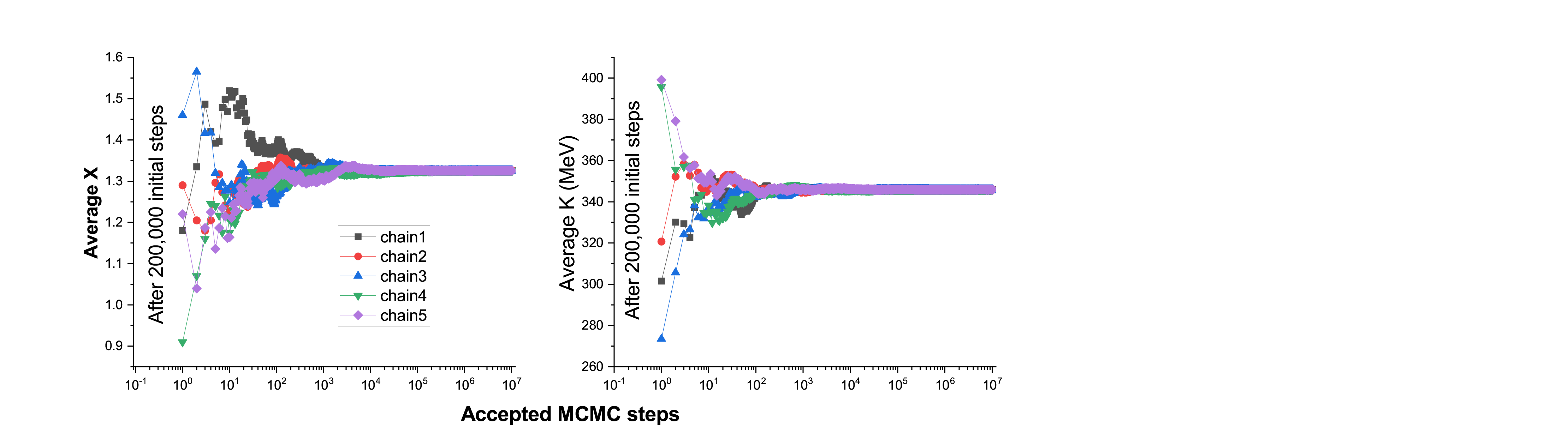}
\caption{Running averages of the X (left) and K (right) parameters as functions of the number of accepted MCMC steps after 200,000 initial steps in 5 independent chains. }\label{steps}
\end{figure}

In our MCMC sampling of posterior PDFs we use the Metropolis-Hastings algorithm. The running averages of model parameters are normally used to check if and when the MCMC has reached equilibrium. Shown in Fig. \ref{steps} are the running averages of X (left) and K (right) parameters as functions of the number of accepted MCMC steps after 200,000 initial steps in 5 independent MCMC chains. It is seen that it took about an additional $10^5$ steps for each chain to reach equilibrium. It is also interesting to note that while the 5 independent chains started at very different initial state (X and K values), they all reach the same equilibrium state as they should. Moreover, the initial fluctuations of X and K are anti-correlated as one expects. Technically, we notice that the number of steps necessary for the MCMC to reach equilibrium strongly depend on the prior ranges we use for X and K. Starting from narrow ranges if our prior knowledge permits it, the chains will reach equilibrium stages much faster. For the purposes of this work, we use the broad prior ranges for X and K given earlier.

\begin{figure*}[thb]
\centering
\includegraphics[width=1.2\textwidth]{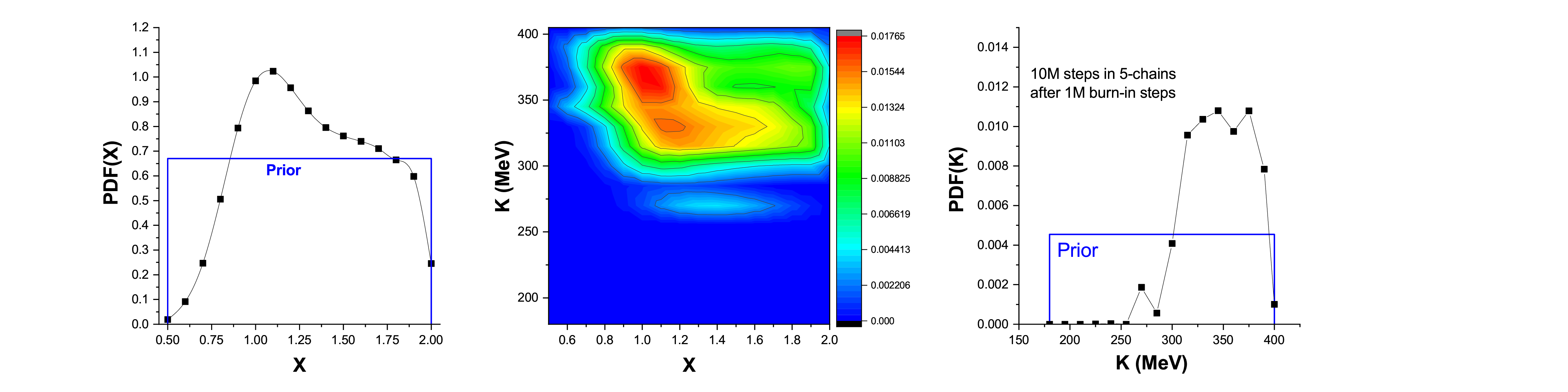}
\caption{Posterior PDFs of X (left) and K (right) as well as their correlation (middle) using 50 million steps from combining the 5 independent chains shown in Fig. \ref{steps} after throwing away the first 1 million steps from each chain. }\label{pdfs}
\end{figure*}
Shown in Fig. \ref{pdfs} are the posterior PDFs of X (left) and K (right) as well as their correlation (middle) using 50 million steps from combining the 5 independent chains after throwing away the first 1 million steps from each chain.  For comparisons, the prior PDFs are also shown.

It is seen that the most probable value (MPV) of X and K are 1.1 and 375 MeV, respectively. Moreover, the PDFs of both X and K are asymmetric and they are anti-correlated as shown in the middle panel. Most interestingly, MPV (X)=1.1 and the highly asymmetric posterior PDF of X is preferentially in the region of $X\ge 1$. Together they clearly provide circumstantial evidence for an enhanced in-medium baryon-baryon cross section.
Since the posterior PDFs of both X and K are asymmetric, following the procedure given in Ref. \cite{Turkkan14} we calculate the highest posterior density (HPD) interval at 68\% confidence level according to
\begin{equation}\label{HPD}
  \int_{p_{i\mathrm{L}}}^{p_{i\mathrm{U}}}\mathrm{PDF}(p_{i})dp_{i}=0.68
\end{equation}
where $p_{i\mathrm{L}}$ ($p_{i\mathrm{U}}$) is the lower (upper) limit of the corresponding HPD interval of the parameter $p_i$ (X or K).
We found that the 68\% confidence HPD intervals for X and K are $0.92\sim1.60$ and $315\sim 375$ MeV, respectively. Since the posterior PDFs of both X and K are highly asymmetric, it is also useful to know that their mean values are 1.32 and 346 MeV, respectively. Since the MPVs are shown visually in their posterior PDFs, we find it is more informative to present the HPD intervals at 68\% confidence levels with respect to the mean values of X and K. Thus, the HADES data clearly prefers an enhanced in-medium cross section with a mean $X=1.32^{+0.28}_{-0.40}$ and a stiff EOS with $K=346^{+29}_{-31}$ MeV, respectively, at 68\% confidence level.

\section{Effects of the prior ranges}

\begin{table}
\centering
\caption{10 training data points on the Latin Hyperlattice with $X$ between 0.5-2.0 and $K$ between 400 and 450 (MeV). The slope $F_1$ of $v_1$ at mid-rapidity and $v_2$ are for free protons}
\begin{tabular}{cccc}\label{10training}
\\\hline
X & K (MeV) & $F_1$ & $v_2$ \\\hline
0.532, &443.87, &0.244E+00, &-0.281E-01\\
0.789, &400.08, &0.366E+00, &-0.405E-01  \\
0.883, &420.80, &0.405E+00, &-0.556E-01 \\
0.969, &408.13, &0.441E+00, &-0.524E-01  \\
1.178, &433.74, &0.498E+00, &-0.723E-01 \\
1.378, &418.28, &0.528E+00, &-0.820E-01  \\
1.460, &429.53, &0.558E+00, &-0.849E-01  \\
1.659, &435.39, &0.598E+00, &-0.945E-01  \\
1.790, &445.35, &0.616E+00, &-0.976E-01 \\
1.899, &411.82, &0.595E+00, &-0.966E-01  \\
\hline
\end{tabular}
\end{table}
\begin{figure}[thb]
\centering
\includegraphics[width=1.5\textwidth]{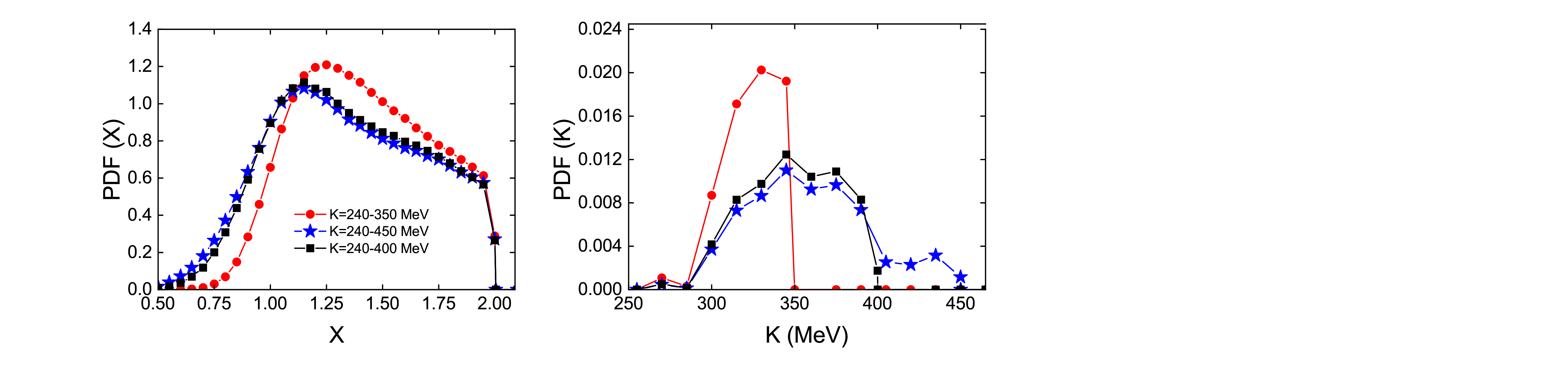}
\caption{Comparisons of PDFs for X (left) and K (right) parameters obtained with the three different prior ranges of K but the same prior range of X between 0.5 and 2.0 }\label{comp}
\end{figure}
\begin{table}
\centering
\caption{Posterior means of $X$ and $K$ as well as their MPVs with $68\%$ HPD boundaries for the three prior ranges for K all with the prior range of $X$ between 0.5 and 2.}
\begin{tabular}{ccccc}\label{post}
\\
\hline
Prior range of K (MeV) & Mean X & Mean K(MeV) & MPV of X & MPV of K(MeV) \\\hline
240-350 &1.38, &325.38, & $1.25^{+0.40}_{-0.15}$, &$330^{+15}_{-15}$ \\
240-400 &1.32, &349.98, & $1.20^{+0.45}_{-0.15}$, &$375^{+15}_{-45}$\\
240-450 &1.31, &358.97, & $1.15^{+0.50}_{-0.20}$, &$375^{+15}_{-45}$\\
\hline
\end{tabular}
\end{table}

According to the Bayesian theorem of Eq. (\ref{Bay1}), the posterior PDFs are proportional to the prior PDFs. In the above analyses, we have used uniform priors in the range of X between 0.5 and 2.0 and $K$ between 180 and 400 MeV.
The prior ranges are educated guesses based on our reviews of the literature as outlined in the introduction section. There is no physical reason that these priors is precise and one of the purposes of Bayesian analysis is to improve them based on the posteriors obtained. One quick lesson we can learn from inspecting the PDFs in Fig.\ \ref{pdfs} is that the lower limit 180 MeV for K used in its prior is too small although such a lower limit was used previously in studying particle production in heavy-ion reactions at GSI energies. Nevertheless, such choice only leads to the waste of some computing time. On the other hand, our choice for the upper limit of 2 for the prior of X parameter obviously leads to the cutoff of its posterior PDF at X=2.
What is the effect of the upper limit chosen for the prior of the K parameter? In the above studies, we have chosen 400 MeV as the maximum value of K based on the current knowledge of the community as discussed earlier. It is not clear if the sharp drop of the PDF of K near K=400 MeV shown in the right panel of Fig.\ \ref{pdfs} is due to the cutoff by the prior PDF or  requirement of the HADES data itself through the likelihood function. To clarify this issue, we performed three new Bayesian calculations all with the same prior range for X between 0.5 and 2.0 as before, and the same lower limit of 240 MeV but different upper limits for the prior K : $K_{max}$=350 MeV, $K_{max}$=400 MeV, and $K_{max}$=450 MeV, respectively.
For the first two cases, we use the available training data set prepared with K between 180 MeV and 400 MeV. For the last case with $K_{max}$=450 MeV, we added 10 additional training points prepared on the Latin Hyperlattice with K between 400 MeV and 450 MeV as listed in Table \ref{10training}. Compared to the HADES data of $F_1=0.46\pm 0.03$ and $v_2=-0.06\pm 0.01$, while most of the new results with large K values are far away from the $1\sigma$ boundaries of the HADES data, some combinations of small X with large K (e.g. with $X=0.883~ \& ~K=420.08$ MeV leading to $F_1=0.405 ~\& ~v_2=-0.056$) are within the $1\sigma$ boundaries of the data. Thus, the K is allowed to go above 400 MeV, the question is how big the PDF there may be. We also notice that in the rarely any training point is exactly on the prior boundary.
Thus, the very last point of the posterior PDF is always forced to go to zero totally by the prior PDF.

Shown in Fig.\ \ref{comp} are comparisons of the PDFs of X and K with the three different prior ranges for K. Several interesting observations can be made: (1) limiting the $K_{max}$ to smaller values requires generally large X values as one expects. (2) The sharp drop in PDF of K near $K_{max}=350$ MeV is completely due to the cut-off of its prior PDF as its posterior of PDF stays rather high when its upper boundary is extended to 400 MeV, indicating that the likelihood function itself is still quite high with $K\simeq 350$ MeV. As the $K_{max}$ is increased from 400 to 450 MeV, however,  the decrease of the posterior PDF is mainly due to the nature drop of the likelihood function itself as indicated by its small albeit finite values between $400<K< 450$ MeV. As noticed earlier, the GP is rarely trained on the boundary. It always gives non-zero small finite probabilities when it is asked to extrapolate to parameter spaces beyond the ranges in which it was trained as Gaussian functions only approach zero when the variables are infinitely away from their central values. (3) In all three cases, the posterior PDFs of X are finite as X approaches its prior upper limit 2.0 and their drop to zero is completely due to the prior PDF of X. Will the posterior PDFs of X and K ever become completely independent of their prior ranges? Based on the results presented above, one may expect a positive answer when the $X_{max}$ and $K_{max}$ are extended to around 5 and 500 MeV, respectively, using the same data set and assuming no other physics constraint will come in. Currently we cannot afford such calculations because of the computationally extremely expensive cost in training the GP in such large parameter space and the limited gain qualitatively compared to what we have already done in understanding the physics issues discussed in this work.

To be more quantitative, shown in Table \ref{post} are comparisons of the mean and MPVs of X and K from the above three new calculations. Interestingly, despite of the dependence of the poster PDFs on the prior ranges of the X and K parameters, both the mean and MPVs of X are significantly larger than one and agree well within $1\sigma$ boundaries in all the cases considered. Thus, our qualitative conclusion that the HADES data provides circumstantial evidence for enhanced baryon-baryon scattering cross sections in hot and dense nuclear matter is robust, albeit the quantitative enhancement depends somewhat on our precise knowledge about the nuclear incompressibility.
\\

\section*{Conclusions}
In conclusion, we found circumstantial evidence for enhanced baryon-baryon scattering cross sections in hot and dense nuclear matter compared to their free-space values from Bayesian analyses of the HADES proton flow data using a GP emulator for the IBUU transport model simulator of heavy-ion reactions at intermediate energies. Quantitatively, the mean value of the in-medium baryon-baryon scattering cross section modification factor X is found to be $X=1.32^{+0.28}_{-0.40}$ at 68\% confidence level assuming the nuclear incompressibility K will not exceed 400 MeV.

\section*{Acknowledgement}
BAL is supported in part by the U.S. Department of Energy, Office of Science,
under Award No. DE-SC0013702, the CUSTIPEN (China-
U.S. Theory Institute for Physics with Exotic Nuclei) under
US Department of Energy Grant No. DE-SC0009971. WJX is supported in part by the Shanxi Provincial Foundation for Returned Overseas Scholars under Grant No 20220037, the Natural Science Foundation of Shanxi Province under Grant No 20210302123085, and the Natural Science Foundation of Yuncheng city.
BAL also acknowledges very fruitful discussions with many participants of the INT Programs: (1) INT-22-1 on ``Machine Learning for Nuclear Theory", (2) INT-22-84W on ``Dense Nuclear Matter Equation of State from Heavy-Ion Collisions", (3) INT-23-1a on ``Intersection of Nuclear Structure and High‐energy Nuclear Collisions" and the Institute for Nuclear Theory at the University of Washington in Seattle for its hospitality.




\begin{thebibliography}{99}

\bibitem{Sto86}H. St\"ocker and W. Greiner, Phys. Rep. {\bf 137} (1986) 277.

\bibitem{Bertsch} G.F. Bertsch and S. Das Gupta, Phys. Rep. {\bf 160} (1988) 189.

\bibitem{Cas90} W. Cassing, V.  Metag, U. Mosel and K. Niita, Phys. Rep. {\bf 188} (1990) 363.

\bibitem{das93}S. Das Gupta and G.D. Westfall, Physics Today, {\bf 46}(5) (1993) 34.

\bibitem{res97}W. Reisdorf and H.G. Ritter, Ann. Rev. Nucl. Part. Sci. {\bf 47} (1997) 663.

\bibitem{Bass} S.A. Bass \textit{et al}, Prog. Part. Nucl. Phys. \textbf{41} (1998) 255.

\bibitem{Pawel02}P. Danielewicz, R. Lacey, W. G. Lynch, Science 298 (2002) 1592.

\bibitem{Buss}
O.~Buss, T.~Gaitanos, K.~Gallmeister, H.~van Hees, M.~Kaskulov, O.~Lalakulich, A.~B.~Larionov, T.~Leitner, J.~Weil and U.~Mosel,
Phys. Rept. \textbf{512} (2012) 1.
\bibitem{Heinz13}
U.~Heinz and R.~Snellings,
Ann. Rev. Nucl. Part. Sci. \textbf{63} (2013) 123.

\bibitem{QCD}Xiaofeng Luo, Qun Wang, Nu Xu, Pengfei Zhuang (Eds), ``Properties of QCD Matter at High Baryon Density", Springer Nature Singapore, Singapore, pages 183--285, 2022.
\url{https://link.springer.com/chapter/10.1007/978-981-19-4441-3_4}

\bibitem{EOSBook}
N.~Xu, J.~Stroth, T.~Galatyuk, Y.~Leifels, F.~Q.~Wang, H.~Sako, B.~Hong, X.~Dong, R.~X.~Xu and L.~W.~Chen, \textit{et al.}, ``Nuclear Matter at High Density and Equation of State,'' Chapter 4 in the Book "Properties of QCD Matter at High Baryon Density", Springer Nature Singapore, Singapore, pages 183--285, 2022.
\url{https://link.springer.com/chapter/10.1007/978-981-19-4441-3_4}

\bibitem{LRP1}
A.~Sorensen, \textit{et al.}
[arXiv:2301.13253 [nucl-th]].

\bibitem{LRP2}
A.~Lovato, \textit{et al.}
[arXiv:2211.02224 [nucl-th]].

\bibitem{LRP3}
P.~Achenbach, \textit{et al.}
[arXiv:2303.02579 [hep-ph]].

\bibitem{Bertsch:1988xu}
G.~F.~Bertsch, G.~E.~Brown, V.~Koch and B.~A.~Li,
Nucl. Phys. A \textbf{490} (1988) 745.

\bibitem{Xu:1991zz}
H.~M.~Xu,
Phys. Rev. Lett. \textbf{67} (1991) 2769.

\bibitem{Westfall:1993zz}
G.~D.~Westfall,\textit{et al.}
Phys. Rev. Lett. \textbf{71} (1993) 1986.

\bibitem{Klakow:1993dj}
D.~Klakow, G.~Welke and W.~Bauer,
Phys. Rev. C \textbf{48} (1993) 1982.

\bibitem{TLi93}
T.~Li, \textit{et al.}
Phys. Rev. Lett. \textbf{70} (1993) 1924.

\bibitem{Alm:1995chb}
T.~Alm, G.~R\"opke, W.~Bauer, F.~Daffin and M.~Schmidt,
Nucl. Phys. A \textbf{587} (1995) 815.

\bibitem{BALI1}
B.~A.~Li, C.~M.~Ko and W.~Bauer,
Int. J. Mod. Phys. E \textbf{7} (1998) 147.

\bibitem{Zheng99}
Y.~M.~Zheng, C.~M.~Ko, B.~A.~Li and B.~Zhang,
Phys. Rev. Lett. \textbf{83} (1999) 2534.

\bibitem{LiSustich}
B.~A.~Li and A.~T.~Sustich,
Phys. Rev. Lett. \textbf{82} (1999) 5004.


\bibitem{Danielewicz:2002he}
P.~Danielewicz,
Acta Phys. Polon. B \textbf{33} (2002) 45.

\bibitem{BALI2}
B.~A.~Li, B.~J.~Cai, L.~W.~Chen and J.~Xu,
Prog. Part. Nucl. Phys. \textbf{99} (2018) 29.

\bibitem{Herman1}T.~Gaitanos, C.~Fuchs and H.~H.~Wolter,
Phys. Lett. B \textbf{609} (2005) 241.

\bibitem{Zhang:2007gd}
Y.~Zhang, Z.~Li and P.~Danielewicz,
Phys. Rev. C \textbf{75} (2007) 034615.

\bibitem{PLi18}
P.~Li, Y.~Wang, Q.~Li, C.~Guo and H.~Zhang,
Phys. Rev. C \textbf{97} (2018) 044620.

\bibitem{Li:2022wvu}
P.~Li, Y.~Wang, Q.~Li and H.~Zhang,
Phys. Lett. B \textbf{828} (2022) 137019.

\bibitem{Haar}
B.~Ter Haar and R.~Malfliet,
Phys. Rev. C \textbf{36} (1987) 1611.

\bibitem{Gale}D. Persram and C. Gale, Phys. Rev. C{\bf 65} (2002) 064611.

\bibitem{LiChen05} B. A. Li and L. W. Chen, Phys. Rev. C {\bf 72} (2005) 064611.

\bibitem{Li:2005iba}
B.~A.~Li, P.~Danielewicz and W.~G.~Lynch,
Phys. Rev. C \textbf{71} (2005) 054603.

\bibitem{Li:1993rwa}
G.~Q.~Li and R.~Machleidt,
Phys. Rev. C \textbf{48} (1993) 1702.

\bibitem{Lom96}
G.~Giansiracusa, U.~Lombardo and N.~Sandulescu,
Phys. Rev. C \textbf{53} (1996) R1478.

\bibitem{Fuchs01}
C.~Fuchs, A.~Faessler and M.~El-Shabshiry,
Phys. Rev. C \textbf{64} (2001) 024003.

\bibitem{Sa}
L.~White and F.~Sammarruca,
Phys. Rev. C \textbf{90} (2014) 044607.

\bibitem{Carlos}
B.~Chen, F.~Sammarruca and C.~A.~Bertulani,
Phys. Rev. C \textbf{87} (2013) 054616.

\bibitem{BALI}B. A. Li, L. W. Chen and C. M. Ko, Phys. Rep. {\bf 464} (2008) 113.


\bibitem{Herman}H.~Wolter \emph{et al}. [TMEP], Prog. Part. Nucl. Phys. {\bf 125} (2022) 103962.

\bibitem{Hades}J. Adamczewski-Musch et al (HADES Collaboration), Phys. Rev. Lett. {\bf 125} (2020) 262301.

\bibitem{Ha2}J.~Adamczewski-Musch \textit{et al.} [HADES],
Eur. Phys. J. A \textbf{59} (2023) 80.

\bibitem{libauer2} B.A. Li, W. Bauer and G.F. Bertsch, Phys. Rev. C{\bf 44} (1991) 2095.

\bibitem{LiBA04} B. A. Li, C. B. Das, S. Das Gupta and C. Gale, Phys. Rev. C {\bf 69} (2004) 011603;
{\it ibid}, Nucl. Phys. A {\bf 735} (2004) 563.


\bibitem{Xie20}
W.~J.~Xie and B.~A.~Li,
J. Phys. G \textbf{48} (2021) 025110.

\bibitem{Fuchs}C. Fuchs, \textit{Prog. Part. Nucl. Phys. }\textbf{56} (2006) 1.

\bibitem{Lynch09}W.G. Lynch W G \textit{et al}, \textit{Prog. Part. Nucl. Phys.} \textbf{62} (2009) 427.

\bibitem{Aich87}
J.~Aichelin, A.~Rosenhauer, G.~Peilert, H.~Stoecker and W.~Greiner,
Phys. Rev. Lett. \textbf{58} (1987) 1926.

\bibitem{Gale87}C. Gale, G. Bertsch, and S. Das Gupta, Phys. Rev. C 35 (1987) 1666.

\bibitem{Prakash}M. Prakash, T. T. S. Kuo, and S. Das Gupta,
Phys. Rev. C {\bf 37} (1988) 2253.

\bibitem{Gerd}G. M. Welke, M. Prakash, T. T. S. Kuo, S. Das Gupta, and C. Gale,
Phys. Rev. C {\bf 38} (1988) 2101.

\bibitem{Pan-Pawel} Qiubao Pan and Pawel Danielewicz, Phys. Rev. Lett. {\bf 70} (1993) 2062; Erratum Phys. Rev. Lett. {\bf 70} (1993) 3523.

\bibitem{Cozma}M.D. Cozma, Euro Phys. J. A {\bf 54} (2018) 40.

\bibitem{Fuchs06}
C.~Fuchs and H.~H.~Wolter,
Eur. Phys. J. A \textbf{30} (2006) 5.

\bibitem{Li97}
B.~A.~Li, C.~M.~Ko and Z.~Z.~Ren,
Phys. Rev. Lett. \textbf{78} (1997) 1644.

\bibitem{Linpa}
B.~A.~Li,
Nucl. Phys. A \textbf{708} (2002) 365.
\bibitem{Ste}
J.~Steinheimer, A.~Motornenko, A.~Sorensen, Y.~Nara, V.~Koch and M.~Bleicher,
Eur. Phys. J. C \textbf{82} (2022) 911.

\bibitem{Rei}
T.~Reichert, O.~Savchuk, A.~Kittiratpattana, P.~Li, J.~Steinheimer, M.~Gorenstein and M.~Bleicher,
Phys. Lett. B \textbf{841} (2023) 137947.
\bibitem{Oma}
M.~Omana Kuttan, J.~Steinheimer, K.~Zhou and H.~St\"ocker,
[arXiv:2211.11670 [nucl-th]].
\bibitem{Oli}
D.~Oliinychenko, A.~Sorensen, V.~Koch and L.~McLerran,
[arXiv:2208.11996 [nucl-th]].

\bibitem{pawel85}P. Danielewicz and G. Odyniec, Phys. Lett. {\bf B157} (1985) 146.

\bibitem{oll}J.~Y.~Ollitrault,
Phys. Rev. D \textbf{46} (1992) 229.

\bibitem{art}A. Poskanzer and S.A. Voloshin, Phys. Rev. C{\bf 55} (1998) 1671.


\bibitem{FOPI}
W.~Reisdorf \textit{et al.} [FOPI],
Nucl. Phys. A \textbf{876} (2012) 1.

\bibitem{Ha3}
B.~Kardan [HADES],
PoS \textbf{CPOD2017} (2018) 049.

\bibitem{GP}C.E. Rasmussen, and C.K. I. Williams, (2006). Gaussian Processes for Machine Learning. Cambridge,
MA: MIT Press. \url{http://gaussianprocess.org/gpml/}

\bibitem{LH}M.D. McKay, R.J. Beckman, W.J. Conover, ``A Comparison of Three Methods for Selecting Values of Input Variables in the Analysis of Output from a Computer Code". Technometrics. American Statistical Association. 21 (2): 239–245

\bibitem{Li-Jake}
B.~A.~Li and J.~Richter,
 Nucl. Phys. {\bf A 1034} (2023) 122640.

\bibitem{Pratt}
S.~Pratt, E.~Sangaline, P.~Sorensen and H.~Wang,
Phys. Rev. Lett. \textbf{114} (2015) 202301.

\bibitem{Scott}
J.~Novak, K.~Novak, S.~Pratt, J.~Vredevoogd, C.~Coleman-Smith and R.~Wolpert,
Phys. Rev. C \textbf{89} (2014) 034917.

\bibitem{Bass2}
J.~E.~Bernhard, P.~W.~Marcy, C.~E.~Coleman-Smith, S.~Huzurbazar, R.~L.~Wolpert and S.~A.~Bass,
Phys. Rev. C \textbf{91} (2015) 054910.

\bibitem{Bass3}
J.~E.~Bernhard, J.~S.~Moreland and S.~A.~Bass,
Nature Phys. \textbf{15} (2019) 1113.

\bibitem{Ohio}
D.~R.~Phillips, R.~J.~Furnstahl, U.~Heinz, T.~Maiti, W.~Nazarewicz, F.~M.~Nunes, M.~Plumlee, M.~T.~Pratola, S.~Pratt and F.~G.~Viens, \textit{et al.}
J. Phys. G \textbf{48} (2021) 072001.

\bibitem{Turkkan14} N. Turkkan and T. Pham-Gia, \textit{J. Stat. Comput. Sim.} \textbf{44} (2014) 243.


\end{thebibliography}
\end{document}